**Grzegorz KOCZAN, Paweł KOZAKIEWICZ**

# UNIFIED ANISOTROPIC STRENGTH CRITERION RANK TWO FOR FIBROUS MATERIALS LIKE WOOD

*Wood is a fibrous orthotropic material additionally characterized with sign--sensitivity. Thus, determining the universal conditions of its strength constitutes a complex task.*
*In their work, the authors present the anisotropic generalization of the Huber criterion which is closer to Norris's proposal, as opposed to Hill (and Hofmann) and Mises proposal. The obtained criterion incorporates conditions which are additionally imposed on a special case of Tsai-Wu criterion for composite materials.*

## Introduction

Wood as an anisotropic and sign-sensitive material is difficult to describe mechanically. The state of stress is described with the stress tensor of $\sigma_{ij} = \sigma_{ji}$, whose off-diagonal components are shearing stresses $\tau_{ij} = \sigma_{ij}$ for $i \neq j$ ($\tau_{ii} = 0$) in the classical sense. The indexes which occur here, have the values of 1, 2 and 3, and in the case of wood they can be connected with the anatomic (longitudinal L, radial R and tangential T) directions respectively. If at least two independent components of stress are non-zero, the subject of the effort of the material (wood) arises. The answer to this question is the strength criteria or failure criteria. These criteria can be divided into isotropic and anisotropic.

Grzegorz KOCZAN✉ (*gkoczan@fuw.edu.pl*), Paweł KOZAKIEWICZ (*pawel_kozakiewicz@sggw.pl*), Departament of Wood Technology, SGGW, Warsaw, Poland

The two most essential isotropic criteria refer to shear stresses. The measure of material effort in the Coulomb-Tresca criterion is twice the maximum value of the shear stress [Tresca 1864][1]. For a flat state of compression, this criterion equals the following inequality:

$$(\sigma_{11}-\sigma_{22})^2+4\tau_{12}^2\leqslant r^2 \tag{1a}$$

$$|(\sigma_{11}+\sigma_{22})/2\pm\sqrt{(\sigma_{11}-\sigma_{22})^2/4+\tau_{12}^2}|\leqslant r \tag{1b}$$

where: $r$ – strength of material. The second condition (1b) applies when, $\sigma_{11}\sigma_{22}\geqslant\tau_{12}^2$, which means that the maximum shear stress exceeds the initial plane considered here. This nonanalytical condition (1b) can also be presented as the following:

$$(\sigma_{11}+\sigma_{22})^2/4\leqslant r^2, \qquad \tau_{12}^2-\sigma_{11}\sigma_{22}\pm r(\sigma_{11}+\sigma_{22})\leqslant r^2 \tag{1b.1}$$

An interesting thing here is the occurrence of the first rank expressions in stresses[2].

The second isotropic criterion refers to the work of shear stresses. Huber's criterion [Huber 1904] stipulates that the measure of the material's effort in a complex state is a value of normal stress which gives the same distortion strain energy as the stress state. Therefore:

$$\frac{(\sigma_{22}-\sigma_{33})^2+(\sigma_{33}-\sigma_{11})^2+(\sigma_{11}-\sigma_{22})^2}{2}+3(\tau_{23}^2+\tau_{31}^2+\tau_{12}^2)\leqslant r^2 \tag{2}$$

or equivalently:

$$\sigma_{11}^2+\sigma_{22}^2+\sigma_{33}^2-\sigma_{22}\sigma_{33}-\sigma_{33}\sigma_{11}-\sigma_{11}\sigma_{22}+3(\tau_{23}^2+\tau_{31}^2+\tau_{12}^2)\leqslant r^2 \tag{2.1}$$

The first to use the distortion strain energy and to publish this criterion was a renowned Polish engineer Maximilian Huber [Timoshenko 1953][3]. Huber is also famous for his theory of steel-reinforced concrete, which he called orthotropic [Huber 1921; Huber 1929] and what made it similar to wood.

Nine years after the formation of the criterion by Huber it was independently confirmed by von Mises[4]. Mises also noticed that it can be described using a second invariant $J_2$ of the deviator stress tensor in the following form:

$$3J_2(\sigma_{ij})\leqslant r^2 \tag{2.2}$$

Where the left side equals the left side in (2) or (2.1).

[1]It is not about the maximum value of $\tau_{12}$ but the maximum value of the largest $\tau'_{i'j'}$ component of all the coordinate systems.

[2]The ± sign makes the criterion symmetrical for both compression and tension.

[3]The same idea was proposed by Maxwell in 1856, in a letter to Thomson. This letter was published in 1936 [e.g. Kordzikowski 2012].

[4]In 1924 it was also confirmed by Hencky (20 years after Huber).

An interesting criterion based on (non-deviatoric) main invariants $I_1$, $I_2$, $I_3$ was proposed by Matsuoka and Nakai [1985][5]. This work does not encompass the third rank criteria however, so it is not going to be discussed here.

The criteria (1) and (2) makes them applicable to elastic-plastic isotropic materials, but wood is anisotropic. The first ever anisotropic criterion was introduced by von Mises [1928]:

$$A\sigma_{11}^2+B\sigma_{22}^2+C\sigma_{33}^2-F\sigma_{22}\sigma_{33}-G\sigma_{33}\sigma_{11}-H\sigma_{11}\sigma_{22}+ \\ +L\tau_{23}^2+M\tau_{31}^2+N\tau_{12}^2\leqslant 1 \tag{3}$$

This criterion is well discussed in the presentation by Zahr Vinuela and Perez Castellanos [2015]. It is evident that Mises was generalising the isotropic criterion in version (2.1) and not in version (2), and not in his original version (2.2).

Another route of generalisation chosen by Hill, was when he decreased the number of parameters in his criterion [Hill 1948][6].

$$\frac{F(\sigma_{22}-\sigma_{33})^2+G(\sigma_{33}-\sigma_{11})^2+H(\sigma_{11}-\sigma_{22})^2}{2}+L\tau_{23}^2+M\tau_{31}^2+N\tau_{12}^2\leqslant 1 \tag{4}$$

We can see that Hill was generalising the isotropic criterion in version (2)[7].

Yet another criterion was proposed by Norris 14 years later. It had a structure of three two-dimensional sections [Norris 1962].

$$\frac{\sigma_{11}^2}{X^2}+\frac{\sigma_{22}^2}{Y^2}-\frac{\sigma_{11}\sigma_{22}}{XY}+\frac{\tau_{12}^2}{T^2}\leqslant 1 \tag{5a}$$

$$\frac{\sigma_{22}^2}{Y^2}+\frac{\sigma_{33}^2}{Z^2}-\frac{\sigma_{22}\sigma_{33}}{YZ}+\frac{\tau_{23}^2}{R^2}\leqslant 1 \tag{5b}$$

$$\frac{\sigma_{33}^2}{Z^2}+\frac{\sigma_{11}^2}{X^2}-\frac{\sigma_{33}\sigma_{11}}{ZX}+\frac{\tau_{31}^2}{S^2}\leqslant 1 \tag{5c}$$

Hill's criterion (4) used for monotropic materials[8] in a stress plane parallel to the monotropic axis, has the following form [Azzi and Tsai 1965]:

$$\frac{\sigma_{11}^2}{X^2}+\frac{\sigma_{22}^2}{Y^2}-\frac{\sigma_{11}\sigma_{22}}{X^2}+\frac{\tau_{12}^2}{T^2}\leqslant 1 \tag{6a}$$

And in a stress plane perpendicular to the monotropic axis:

$$\frac{\sigma_{22}^2}{Y^2}+\frac{\sigma_{33}^2}{Y^2}-\left(\frac{2}{Y^2}-\frac{1}{X^2}\right)\sigma_{22}\sigma_{33}+\frac{\tau_{23}^2}{R^2}\leqslant 1, \qquad \frac{1}{R^2}=\frac{4}{Y^2}-\frac{1}{X^2} \tag{6b}$$

[5] However, the authors used signs $J_1$, $J_2$, $J_3$, which are usually ascribed to the deviator.
[6] Hill's criterion contains 6 parameters, and Mises 9 independent parameters.
[7] The coefficients in (4) are twice as high than the original coefficients in Hill's criterion.
[8] Wood is approximately transversely isotropic (monotropic).

The case (6b) was implicitly given by Hill [1948], while case (6a) was explicitly stated by Azzi and Tsai 18 years later. Therefore, the formula (6a) is called the Azzi-Tsai criterion [e.g. Guindos 2014] (or Tsai-Hill criterion [e.g. Kolios and Proia 2012]), and its general monotropic case of the Hill criterion is called the Tsai-Hill criterion [e.g. Camanho 2002]. An example of the application of criterion (6a) for beech plywood can be found in Makowski [2013].

None of these criteria takes into consideration the differences in compression and tension strength[9] (strength sign-sensitivity). The first general and theoretical criterion (6a) containing a solution to this problem was proposed in 1966 [Gol'denblat and Kopnov 1966][10]. A simplified version of this solution was presented by Hoffman when he expanded Hill's criterion by first rank terms [Hoffman 1967]:

$$\frac{F(\sigma_{22}-\sigma_{33})^2+G(\sigma_{33}-\sigma_{11})^2+H(\sigma_{11}-\sigma_{22})^2}{2}+L\tau_{23}^2+M\tau_{31}^2+N\tau_{12}^2+$$
$$+K\sigma_{11}+P\sigma_{22}+Q\sigma_{33}\leqslant 1 \tag{7}$$

Analogically, Tsai and Wu expanded Mises criterion by additionally using a matrix form [Tsai and Wu 1971]:

$$F_i\sigma_{ii}+F_{ij}\sigma_{ii}\sigma_{jj}+\frac{1}{2}F_{\underline{ij}\,\underline{ij}}\tau_{ij}^2\leqslant 1 \tag{8}$$

Where summation convention was used together with superscript $\underline{ij}=9-i-j$ conforming to the Voigt extended cyclic notation for shear stresses[11]. The expanded version of the criterion is:

$$F_1\sigma_{11}+F_2\sigma_{22}+F_3\sigma_{33}+F_{11}\sigma_{11}^2+F_{22}\sigma_2^2+F_{33}\sigma_{33}^2+2F_{23}\sigma_{22}\sigma_{33}+$$
$$+2F_{31}\sigma_{33}\sigma_{11}+2F_{12}\sigma_{11}\sigma_{22}+F_{44}\tau_{23}^2+F_{55}\tau_{31}^2+F_{66}\tau_{12}^2\leqslant 1 \tag{8.1}$$

This criterion contains a certain ambiguity in how it determines interaction coefficients, which are not defined by strengths only in L, R and T directions. Thus, we should additionally measure the strength for biaxial and equibiaxial stresses ($\sigma_{22}=\sigma_{33}$ or $\sigma_{33}=\sigma_{11}$ or $\sigma_{11}=\sigma_{22}$). Another method is to assume the following interaction coefficients:

$$2F_{ij}=-\sqrt{F_{ii}F_{jj}}, \qquad i\neq j \tag{9}$$

It must be emphasized that condition (9) does not result from the criterion formula (8.1), therefore it is regarded here as a separate but not autonomous criterion. It is sometimes called a criterion for closed-cell PVC cellular foam

[9]In a non-analytical approach, separated signs $R_{\parallel}^{\pm}$ , $R_{\perp}^{\pm}$ [e.g. Garab and Szalai 2010].

[10]According to [Kyzioł 2009] their criterion is practically equivalent to Tsai-Wu (8), but it's probably not true.

[11]Strong equality of (8) and (8.1) is our contribution, where $F_{ij}=F_{ji}$, $\tau_{ii}=0$.

[Abrate 2008] or a criterion for composite materials. It is also frequently and wrongly called Tsai-Wu criterion [e.g. Gdoutos and Daniel 2008, Cabrero and Gebremedhin 2010] or more precisely the simplified Tsai-Wu criterion [Feldhusen and Krishnamoorthy 2009]. In terms of interaction, this version of the criterion conforms to Norris's criterion and does not conform to Hill's (and Azzi-Tsai) criterion. We have not found in the literature of the subject a derivation of condition (9), and it is regarded only as an assumption. For instance, in DeTeresa and Larsen [2001] other interaction terms are postulated, and in Cowin [1979], van der Put [1982] and Liu [1984] still others. In a publication on Douglas-fir laminated veneer Clouston et al. [1998] we read: *Despite these efforts (with Hankinson's formula*[12]*) a standard method of determining* $F_{12}$ *was never established*. A thorough discussion about older strength criteria regarding wood was presented in Clouston [1995]. A synthesis of more advanced research on wood also including the criteria of the third rank was presented by van der Put [2015].

## Materials and methods

The first part of the methodology refers to the theory. Its role in solving technical problems was emphasized by one of the pioneers of the strengths hypotheses M. Huber [Huber 1912; Huber 1927]. The second, experimental part of the methodology is focused on a description of the experimental tests aiming at determining the $2F_{12}$-type interaction coefficient.

### Assumptions of mathematical research

It was postulated that the criterion $f(\{\sigma_{ij}\}) \leq 1$ should meet the following assumptions:

i. Be an anisotropic generalisation of Huber's criterion. (Norris's criterion as three two-dimensional formulae does not meet this condition.)
ii. Anisotropic coefficients should stand directly next to the stresses as a measure of directional effort. (This condition is not met by Mises and Hill's criteria, but is met by Norris's criterion).
iii. Asymmetry for compression and tension should be considered (sign--sensitivity), thanks to the first rank terms in stresses (like Gol'denblat--Kopnov, Hoffman, Tsai-Wu).
iv. They should not contain free parameters exceeding beyond the measurements of the main components of stress. (This condition is not met by the general Mises and Tsai-Wu criteria.)

[12]Hankinson's formula does not conform with (9) [Clouston 1995].

v. In general, non-trivial additional conditions should not be needed. In other words, we wish to build an autonomous criterion. (This condition is not met by a simplified Tsai-Wu criterion).
vi. It should be possible to write in an unambiguous and compact index form. (A condition met by the Tsai-Wu criterion after adjustments by the authors).
vii. It should be possible to write it using functions determined by main or deviator invariants. (A condition met only by Huber's criterion).

Essentially, only assumption iv. requires experimental verification.

The parameters found in the formula of the criterion were determined based on directional strength. In the case of normal stress, there are only two conditions necessary for tension and compression, e.g.:

$$f(\sigma_{11}^{\max}=+R_{11}^{+},0,0,0,0,0)=1, \qquad f(\sigma_{11}^{\min}=-R_{11}^{-},0,0,0,0,0,)=1 \qquad (10)$$

where: $R_{11}^{+}$, $R_{11}^{-}$ – tension strength and compression strength parallel to the grain. In the case of shear stresses, single conditions are sufficient, e.g.:

$$f(0,0,0,0,0,\tau_{12}^{\max}=R_{12})=1 \qquad (11)$$

where: $R_{12}$ – shear strength parallel to the grain in the tangential plane.

In the case of monotropy with axle 1, apart from conditions $R_{22}^{\pm}=R_{33}^{\pm}$, $R_{12}=R_{31}=\tau_{31}^{\max}$ the following relation occurs:

$$f(0,\sigma_{22}=R_{23},\sigma_{33}=-R_{23},0,0,0)=1 \qquad (12)$$

where $R_{23}=\tau_{23}^{\max}$ is the shear strength across the monotropy axis[13].

The key role in the differentiation of anisotropic strength criteria is played by biaxial stresses. The strength for biaxial shear state can be described as follows:

$$f(\sigma_{11}=\varepsilon_{11}^{-}R_{11}^{-},\sigma_{22}=\varepsilon_{22}^{-}R_{22}^{-},0,0,0,0)=1 \qquad (13)$$

where $\varepsilon_{11}^{-}<0$, $\varepsilon_{22}^{-}<0$ are the measure of destructive effort in both directions of compression (hence the negative marks). The values of these measures may be different for different strength criteria and they depend first and foremost on the interaction terms like $2F_{12}$. An example of biaxial tensions are the tensions occurring under the head in the bending test [Bodig and Jayne 1993]. In this work it is assumed that in such cases, the strength criterion should allow states of the following efforts:

$$|\varepsilon_{11}^{-}|>90\% \quad \wedge \quad |\varepsilon_{22}^{-}|>60\% \qquad (14)$$

Otherwise, compression would decrease the bending strength by at least 10%, which has not been observed. Analysis of conditions (13) and (14) requires proportions between the strengths for the main wood grain directions.

[13]The above condition arises from Mohr's circle on isotopic plane 23 (RT).

$$\frac{R_{22}^{+}}{R_{22}^{-}} \approx 1, \qquad \frac{R_{11}^{+}}{R_{11}^{-}} = a \approx 2, \qquad \frac{R_{11}^{-}}{R_{22}^{-}} = b \approx 5 \tag{15}$$

Condition (14) and values (15) will be further discussed. Relations (14) may discredit the criteria type (4), (6) and (7) in opposition to the Norris-type criteria (5) and (9). While criteria (3) and (8) do not produce any predictions here in the sense that they do not determine the interaction coefficients.

## Types of experimental tests

The main purpose of the experimental tests was to show the existence of effort states determined by conditions (14) which discredit part of the strength criteria. These states were determined by carrying out crushing tests under a cylindrical head of American sweetgum wood (*Liquidambar styraciflua* L.). The main indentation occurred during a standard destructive bending test (span 24 cm, fig. 1a). Comparative measurements were performed on the undamaged sections of the broken samples. The tests consisted of a perpendicular compression of the wood with the head during non-destructive[14] bending with a small span of the supports (span 10 cm, fig. 1b) and compression with a full bottom support (fig. 1c).

a) 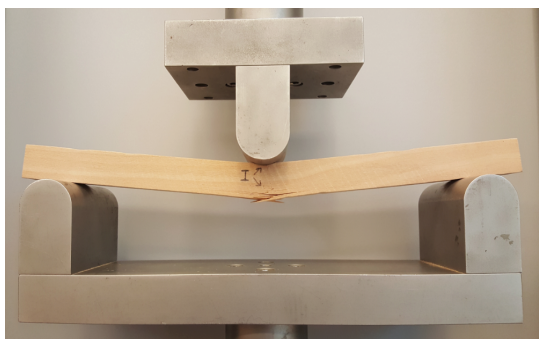 b) 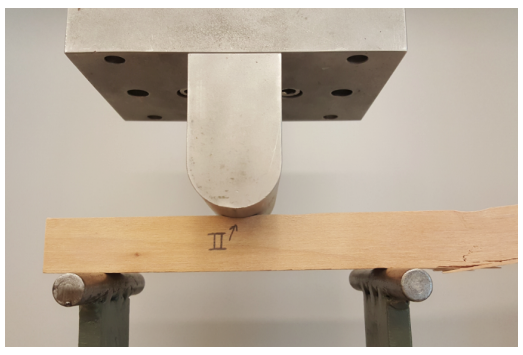 c) 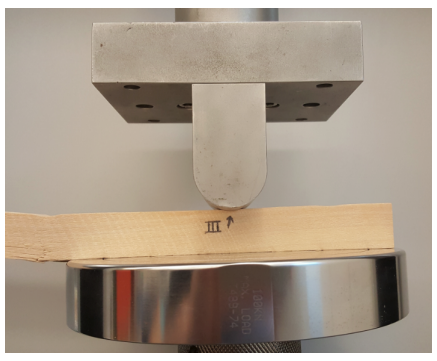

**Fig. 1. Three ways of testing crushing under the cylindrical head: a) static bending (test I), b) short beam compression (test II), c) full bottom support compression perpendicular to the grain (test III)**

In the first set of tests on II and III, the force applied was identical to test I and in the second set of these tests, the force was 4/3 times greater to get a similar indentation. The indentation surface was regarded as a trapezoid whose dimensions were measured with a digital calliper.

The measurement tests performed allowed us to compare the obtained indentations in analogous conditions [Smith et al. 1985][15]. The key difference was that the geometric assumptions for the longitudinal compression stress in test II was about 2.4 times (span ratio) or 1.8 times (3/4 of span ratio) smaller than in test I, and it did not occur in test III at all.

The bending strength was determined by the standard PN-D-04103:1977, which complies with ISO 3133:1975, but is not equivalent to it. The specimens had a size of 20 × 20 × 300 mm with a span length of 240 mm and a radius of the supports of 15 mm. While the norms [ISO 3133:1975; ISO 13061-3:2014] predict a twice greater size of the pressing head (radius 30 mm), which accounts for the main difference between the PN-D-04103:1977 norm which is still used in Poland. The small radius of the pressing head is an additional stimulus for these tests.

[14]In the sense that there is no fracture, but a plastic indentation occurred [Hering et al. 2012].
[15]Other tests such as the standard compression or measurement of hardness would be less relevant.

The samples' moisture content was determined using the stereometric method in accordance to ISO 13061-1:2014 and density using the dried-weighed method in accordance to ISO 13061-2:2014.

## Results and discussion

### Results of mathematical research

The authors' strength criterion (ad. vii.) for wood, based on the invariant $J_2$ for the mean effort deviator and function of the first invariant $I_1$ is as follows:

$$3J_2\left(\frac{\sigma_{ij}}{R_{ij}}\right)+I_1\left(\frac{\sigma_{ii}}{R_i}\right)\leqslant 1 \qquad (16)$$

where the $J_2$ (…) invariant is determined by (2.2) in relation to (2) or (2.1), and function $I_1(\sigma_{ii})= \sigma_{11}+\sigma_{22}+\sigma_{33}$. Parameters $R_{ij}= R_{ji}$ are the measure of mean strengths for all six independent components of stresses in wood's anatomic directions, and $R_i$ is the measure of compression-tension symmetry approaching $\pm\infty$ for the total symmetry ($1/R_i \rightarrow 0$). This criterion is based on mean efforts $\sigma_{ij}/R_{ij}$ in the function of the $J_2$ deviator, and not on the stresses $\sigma_{ij}$ alone. However, the invariant $I_1$ has other (higher) normalizing coefficients $R_i$. The above element and the lack of square in the first rank invariant ($I_1$ versus $I_1^2$) are the two basic differences in criterion (16) from formula "(19)" found in Doyoyo and Wierzbicki [2003][16]. While the prototype version of their formula [Gioux et al. 2000] differs in having $J_2$ raised to the 1/2 power, i.e. square root.

An indicator (ad. vi.), equivalent form of criterion (16) is as follows:

$$\frac{1}{4}\sum_{i,j=1}^{3}\left(\frac{\sigma_{ii}}{R_{ii}}-\frac{\sigma_{jj}}{R_{jj}}\right)^2 + \frac{1}{2}\sum_{i,j=1}^{3}\frac{\tau_{ij}^2}{R_{ij}^2} + \sum_{i=1}^{3}\frac{\sigma_{ii}}{R_i} \leqslant 1 \qquad (16.1)$$

In a partially expanded version, the formula is:

[16]Other differences include the 3D aspect of criterion (16) and the lack of an additional constant $\eta$. Moreover, their criterion "(19)" does not account for sign-sensitivity, and their criterion "(12)" accounts for sign-sensitivity but not full anisotropy (for aluminium).

$$\frac{1}{2}\left(\frac{\sigma_{22}}{R_{22}}-\frac{\sigma_{33}}{R_{33}}\right)^2+\frac{1}{2}\left(\frac{\sigma_{33}}{R_{33}}-\frac{\sigma_{11}}{R_{11}}\right)^2+\frac{1}{2}\left(\frac{\sigma_{11}}{R_{11}}-\frac{\sigma_{22}}{R_{22}}\right)^2+$$
$$+\frac{\tau_{23}^2}{R_{23}^2}+\frac{\tau_{31}^2}{R_{31}^2}+\frac{\tau_{12}^2}{R_{12}^2}+\frac{\sigma_{11}}{R_1}+\frac{\sigma_{22}}{R_2}+\frac{\sigma_{33}}{R_3}\leqslant 1 \tag{16.2}$$

In the fully expanded form, it is:

$$\frac{\sigma_{11}^2}{R_{11}^2}+\frac{\sigma_{22}^2}{R_{22}^2}+\frac{\sigma_{33}^2}{R_{33}^2}-\frac{\sigma_{22}\sigma_{33}}{R_{22}R_{33}}-\frac{\sigma_{33}\sigma_{11}}{R_{33}R_{11}}-\frac{\sigma_{11}\sigma_{22}}{R_{11}R_{22}}+$$
$$+\frac{\tau_{23}^2}{R_{23}^2}+\frac{\tau_{31}^2}{R_{31}^2}+\frac{\tau_{12}^2}{R_{12}^2}+\frac{\sigma_{11}}{R_1}+\frac{\sigma_{22}}{R_2}+\frac{\sigma_{33}}{R_3}\leqslant 1 \tag{16.3}$$

According to (10), the coefficients of this criterion have a simple interpretation:

$$R_{11}=\sqrt{R_{11}^+R_{11}^-},\quad \frac{1}{R_1}=\frac{1}{R_{11}^+}-\frac{1}{R_{11}^-},\quad R_{12}=\tau_{12}^{\max} \tag{17}$$

which is analogous for the other components $R_{ij}$ and $R_i$. Thus, $R_{11}$ can be regarded as the geometric mean of tension and compression strength, and $R_1$ can be treated as a connection in parallel of two resistors, one of which is negative.

In the case of monotropy in (12), an additional condition occurs:

$$R_{23}=\frac{R_{22}}{\sqrt{3}}=\frac{\sqrt{R_{22}^+R_{22}^-}}{\sqrt{3}} \tag{18}$$

The value $\sqrt{3}$ in this relation is the same as for Huber's criterion (2) and Norris's criterion (5b) and is smaller than value 2 resulting from the Coulomb-Tresca criterion (1).

## Results of the experimental test

The shape of the specimen fracture in bending and the accompanying indentation (test I) are presented in figure 2. This picture also presents comparative indentations obtained with the two methods (tests II and III).

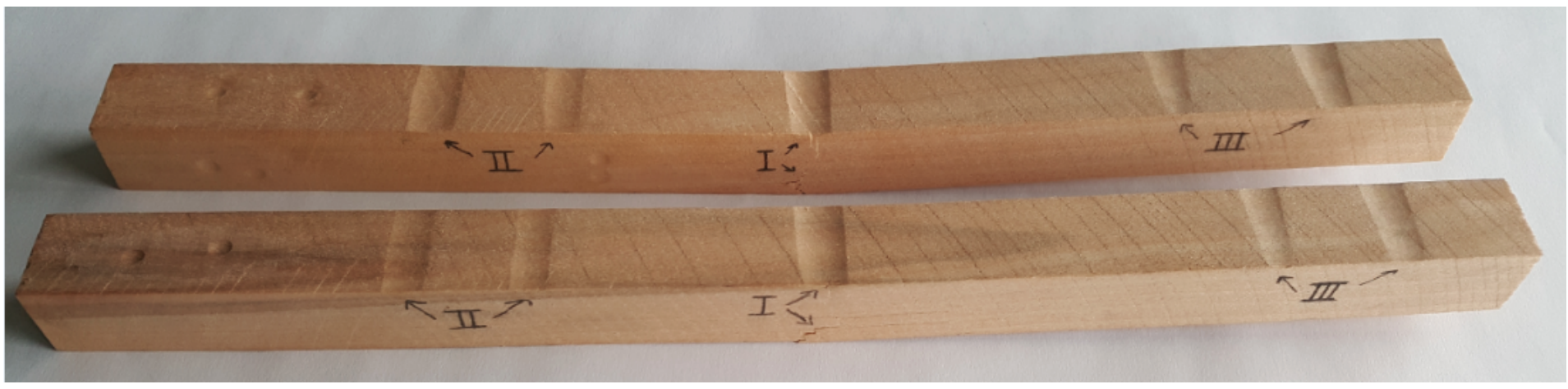

**Fig. 2. A comparison of indentations of American sweetgum wood obtained in: bending test I (in the middle of both beams), short beam compression test II (on the left), full support compression test III (on the right)**

The main results of the three parts of the tests are presented in table 1. They show that the width of the indentations from the breaking test (12 mm) are greater by 1/3 than that of the full support compression test (9 mm). After considering the occurring forces, it means that the average perpendicular stress in bending is 65% of the stress in compression alone. It is worth adding that this is still a lowered value, because the wood indentation in test I was widened due to the curvature of the bending beam[17].

The values from the table are presented in figure 3 against the four strength criteria.

**Table 1. Test results for three methods of compressing American sweetgum wood**

| Moisture content (8.3 ±0.3) %<br>Density (596 ±33) kg/m$^3$<br>Parallel compression (72 ±3) Mpa | Test I<br>Crushing<br>(⊥ , ‖ )<br>in static bending | Test II<br>Compression<br>(⊥ , half ‖)<br>short beam | Test III<br>Compression<br>(⊥ )<br>full support |
|---|---|---|---|
| Number of measurements | 22 | 44 | 44 |
| Used force $F$ [N] | 2809 ±147 | 3195 ±466 | 3335 ±537 |
| Indentation width [mm] | 11.9 ±1.5 | 10.3 ±1.9 | 9.1 ±1.4 |
| Average stress $\sigma_{22}$ [MPa] | -12.1 ±2,0 | -15.8 ±2.5 | -18.6 ±2.5 |
| Average effort $\varepsilon_{22}^{-}$ [%] (relative test III) | -65 ±5 | -85 ±6 | -100 (-95 ±5) |
| Estimate effort $\varepsilon_{11}^{-}$ [%] | ≈ -90 ±10 | -47 ±7 | ≈ 0 ±5 |

The experimental results are presented with the standard deviation as small ellipses and the criteria as large ellipses. Straight segments in the graph shot the critical condition (14).

## Discussion of results

Figure 3 shows that test I allowed us to determine the states which conform to the critical condition (14). At the same time Hill's criterion, which anticipates that $\varepsilon_{11}^{-}=-90\%$, $\varepsilon_{22}^{-}=-54\%$ or $\varepsilon_{11}^{-}=-86\%$, $\varepsilon_{22}^{-}=-60\%$, is inconsistent with this condition. Moreover, the transition from Hill's criterion to Hoffman's criterion, which accounts for the sign-sensitivity of wood, only makes things worse. Norris's criterion meets condition (14) because it assumes a high (even too high) strength at the biaxial compression. The transition from Norris's criterion to our criterion (16), which accounts for sign-sensitivity, moves us closer to the experimental points.

[17] Nevertheless, it is a higher value than that resulting from Hill and Hoffman's criteria.

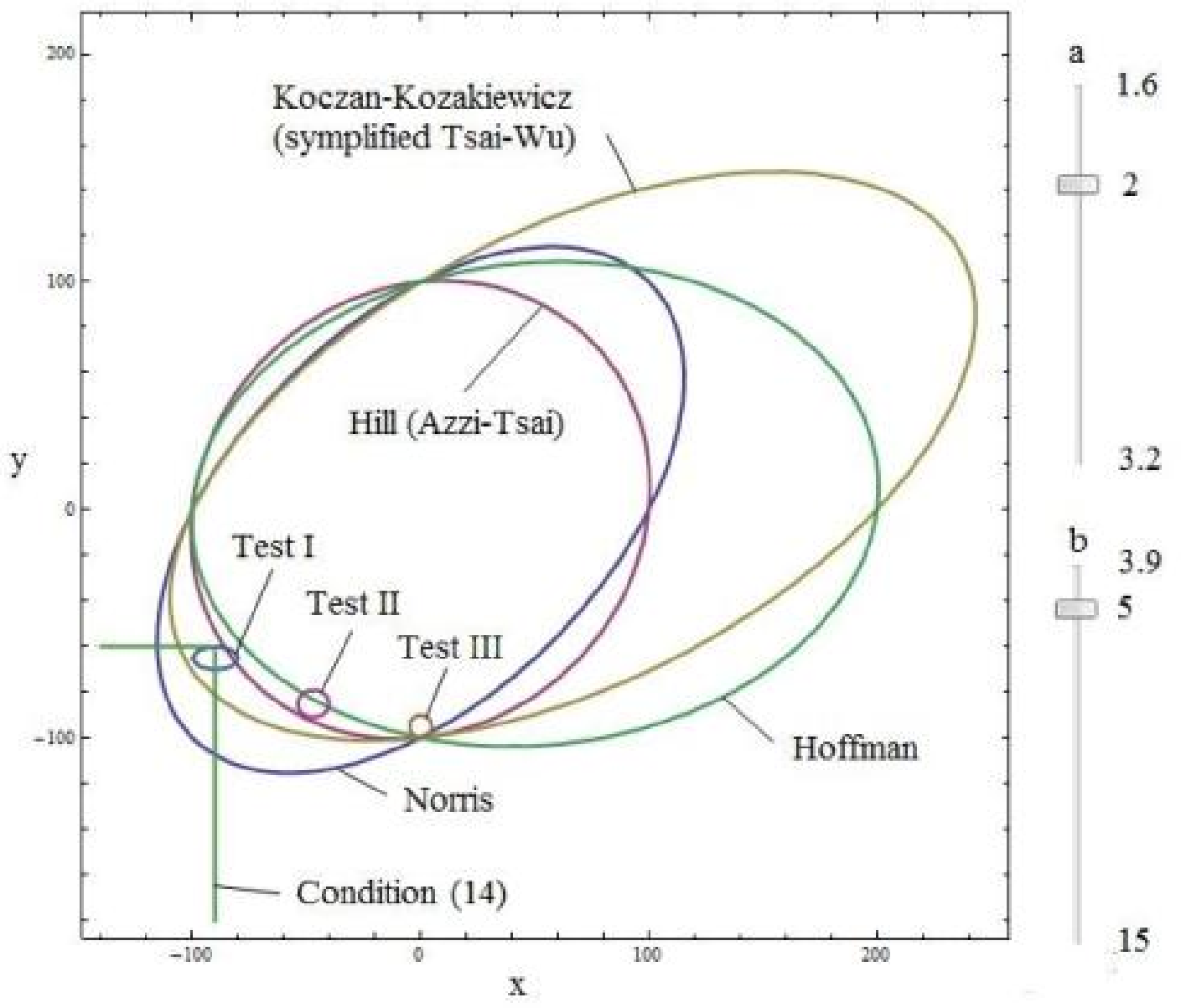

**Fig. 3. The measuring points of the three compression methods against the stress diagrams relating to compression $x=\varepsilon_{11}^{-}$ [%], $y=\varepsilon_{22}^{-}$ [%] in accordance with different strength criteria**

Despite the fact that Hill's criterion is close to the experimental points, the point in test I is almost entirely outside the area or strength of this criterion. Introducing here compression-tension asymmetry (Hoffman's criterion) takes us away from this experimental point, as well as from the area which meets condition (14).

The experimental advantage of Norris's criterion over Hill's criterion was obtained in a work from 1985 for paperboard [Rowlands et al. 1985][18], while it was criticised in van der Put [2015]. The authors of Mascia and Simoni [2013] claim that Hoffman's criterion works the best for two species of Brazilian wood. Thus, it is difficult to determine which criterion is the best only on the basis of experiments and with no theoretical guidelines. However, according to the authors, the bidirectional compression measurements best fit the criterion (16) among those considered.

## Discussion of assumptions about wood

The issue of wood anisotropy is a very complex one and synthetic conclusions require the acceptance of some simplified assumptions. The basic assumption was the strength monotropy of wood, i.e. identifying strength in transverse directions: radial and tangential (about 5% precision for diffuse-porous species). Next, the proportions between the main strengths of the directions of wood were assumed (15). The first equality of compression strength and tension strength perpendicular to the grain is a consequence of the lack of fibres in these directions and is met by wood (including American sweetgum) up to 20% (acc. to Green et al. [1999]). The middle proportion $a \approx 2$ is true for e.g. Norway maple (*Acer platanoides* L.), wych elm (*Ulmus glabra* Huds.) or European larch (*Larix decidua* Mill.) (acc. to Krzysik [1978]). According to the same source, the lowest $a$=1.6 is achieved by hornbeam (*Carpinus betulus* L.) and one of the highest $a$=3.2 is achieved by European birch (*Betula pendula* Roth.). The last proportion $b$≈5 is maintained for white oak (*Quercus alba* L.), true hickory shagbark (*Carya ovata* (Mill.) K. Koch) and sugar maple (*Acer saccharum* Marsh.) (acc. to Green et al. [1999])[19]. According to the same source, one of the lowest values is $b = 4.4$ for hickory pecan nutmeg and the highest is $b = 15$ for black cottonwood (*Populus trichocarpa* Hock.). Here, also $b \approx 10$ for many wood species, including American sweetgum wood. However, in mean stress measurements under the head, this factor was $b = 3.9$. A change in the values of parameter $a$ and parameter $b$ in the discussed spectrum is:

$$1.6 \leqslant a \leqslant 3.2, \quad 3.9 \leqslant b \leqslant 15 \tag{19}$$

[18]The authors consider Norris's criterion to be a Hill criterion-type.

This does not have a considerable effect on the research results, which was tested on the graph in fig. 4 (see sliders).

Now we will analyse more closely condition (14) which is critical for this work. The biaxial state of bidirectional compression (13) occurs during bending tests, directly under the loading head. Already in Koczan and Kozakiewicz [2017], it was noted that Hill's criterion (4) or the Azzi-Tsai criterion (6a) seem to be contradicting the experiment, but Norris's criterion (5) does not. Hill's criterion (Azzi-Tsai's criterion) assumes, e.g.: $\varepsilon_{11}^{-} = -100\%$, $\varepsilon_{22}^{-} = -20\%$ or $\varepsilon_{11}^{-} = \varepsilon_{22}^{-} = -75\%$. While the Norris criterion does not impose such restrictions: $\varepsilon_{11}^{-} = \varepsilon_{22}^{-} = -100\%$ or $\varepsilon_{11}^{-} = -115\%, \varepsilon_{22}^{-} = -58\%$ (just like in the Huber criterion). This means that according to Norris's criterion, compression under the head does not have to influence the bending strength of wood. Hill's criterion does not give that possibility. According to the Baumann graphs of the measurements of stress in bending and according to the Thunell model [Kollmann and Cote 1984], the first value of 90% can be assumed in (14). A tolerance of -10% here, is a gesture towards competitive criteria. Another value for (14) is estimated on the basis of the assumption in accordance with measurements which follow PN-D-04103:1977 that the width contact $w$ of the beam with the head is less than 15 mm (acc. to DIN 52186:1978 it equals the thickness of the beam $h$ = 20 mm):

[19]At the same time the sources Krzysik [1978] and Green et al. [1999] do not contain all the data.

$$|\varepsilon_{22}^{-}| = \frac{b\beta}{18} |\varepsilon_{11}^{-}| \frac{h}{w} \xrightarrow{\beta \approx 2} 67\% \tag{20}$$

where $\beta$ is the ratio of bending and compression strengths parallel to the grain. A tolerance of -7% is another gesture towards competitive criteria.

## Conclusions

We were able to form a strength criterion using invariants (16) or an equivalent index form (16.1) which meet all the theoretical conditions assumed in the methodology i-vii. The advantage of a generalized version of Huber's criterion over Hill's criterion (4) and Hoffman's criterion (7) is comparing (subtraction) wood efforts in different grain directions instead of comparing stresses. The maximum stress parallel to the grain is disproportional to the maximum stress perpendicular to the grain and may be $ab$ = 10 times greater (and even 30 times greater).

This theoretical advantage was initially experimentally verified, using American sweetgum wood as an example, for the state of bidirectional compression under the loading head in bending tests. If criterion (16) turns out to be insufficient for bidirectional tension, it will be necessary to search for more general criteria like Gol'denblat-Kopnov or the third rank criteria [van der Put 2015]. In this situation using Tsai-Wu's criterion, which is marginally more general, will not change much.

The introduced strength criterion (16) or (16.1) is equivalent to the simplified Tsai-Wu criterion (9) together with (8.1). However, the criterion presented in this work is determined by a single autonomous equation (16) or (16.1) which is a direct generalization of Huber's criterion (2). In this sense, it can be regarded as a derivation of the terms of interaction which are incorporated in this deviatoric criterion. Thus, it is a unification of Huber, Norris and the simplified Tsai-Wu criteria which differ from Hill, Azzi-Tsai and Hoffman's criteria. This unification does not include the anisotropic Mises criterion and the general Tsai-Wu criterion, therefore it is a selective unification in the sense of Ockham razor. The synthesis described in this work concerns analytical second rank polynomial criteria.

## List of standards